\documentclass[letter]{aa}  
\usepackage{graphicx}
\usepackage{hyperref}                                                     
\usepackage{color}
\usepackage{txfonts}
\usepackage{placeins}

\newcommand{\Msun}{\,\mathrm{M}_\odot}

\newcommand{\oldcaption}[1]{}

\begin{document}

   \title{Edge-On Disk Study (EODS) II: Thermal Structure of the Flying Saucer Disk}
   
   \author{S. Guilloteau \inst{1}
        \and O. Denis-Alpizar \inst{2}
        \and A. Dutrey \inst{1}
        \and C. Foucher \inst{1}   
        \and S. Gavino \inst{3}
        \and D. Semenov \inst{4,5}
        \and V. Piétu \inst{6} 
        \and E. Chapillon \inst{1,6}  
        \and L. Testi \inst{3}
        \and E. Dartois \inst{7}
        \and E. di Folco \inst{1}
        \and K. Furuya \inst{8}
        \and U. Gorti \inst{9}
        \and N. Grosso \inst{10}
        \and Th. Henning \inst{4}    
        \and J.M. Huré \inst{1} 
        \and A. Kospal \inst{11}
        \and F. LePetit \inst{12}
        \and L. Majumdar \inst{13}
        \and H. Nomura \inst{14}
        \and N.T. Phuong\inst{15}
        \and M. Ruaud \inst{9}
        \and Y.W. Tang \inst{16}
        \and S. Wolf \inst{17} 
        }
        \titlerunning{EODS: Thermal profile of the Flying Saucer disk}
        
   \institute{Univ. Bordeaux, CNRS, Laboratoire d'Astrophysique de Bordeaux (LAB), UMR 5804, F-33600 Pessac, France    \\
              \email{stephane.guilloteau@u-bordeaux.fr} 
    \and Departamento de Física, Facultad de Ciencias, Universidad de Chile, Av. Las Palmeras 3425, Ñuñoa, Santiago, Chile. % Otoniel
    \and 
    Dipartimento di Fisica e Astronomia “Augusto Righi”, ALMA Mater Studiorum - Universiti. Bologna, via Gobetti 93/2, I-40190 Bologna, Italy %Gavino, Testi
    \and Max-Planck-Institut f\"{u}r Astronomie (MPIA), K\"{o}nigstuhl 17, D-69117 Heidelberg, Germany %Semenov
    \and Department of Chemistry, Ludwig-Maximilians-Universit\"{a}t, Butenandtstr. 5-13, D-81377 M\"{u}nchen, Germany  % Semenov
    \and IRAM, 300 Rue de la Piscine, F-38046 Saint Martin d'H\`{e}res, France %Chapillon
    \and Institut des Sciences Moléculaires d'Orsay, CNRS, Univ. Paris-Saclay, Orsay, France %Dartois
    \and RIKEN Cluster for Pioneering Research, 2-1 Hirosawa, Wako-shi, Saitama 351-0198, Japan % Furuya
    \and    Carl Sagan Center, SETI Institute, Mountain View, CA, USA %Gorti
    \and Aix-Marseille Univ, CNRS, CNES, LAM, Marseille, France % Grosso
    \and Konkoly Observatory, Research Centre for Astronomy and Earth Sciences, Konkoly-Thege Miklós út 15-17, 1121 Budapest, Hungary %Kospal
    \and LUX, Observatoire de Paris, PSL Research University,
    CNRS, Sorbonne Universités, 92190 Meudon, France % Le Petit
    \and Exoplanets and Planetary Formation Group, School of Earth and Planetary Sciences, National Institute of Science Education and Research, Jatni 752050, Odisha, India % Mjumbar
    \and National Astronomical Observatory of Japan, Division of Science, 2-21-1 Osawa, Mitaka, Tokyo 181-8588, Kanto Japan %Nomura
    \and Vietnam National Space Center, Vietnam Academy of Science and Technology, 18, Hoang Quoc Viet, Nghia Do, Cau Giay, Ha Noi,Vietnam % Phuong 
    \and Academia Sinica Institute of Astronomy and Astrophysics, 11F of AS/NTU Astronomy-Mathematics Building, No.1, Sec.4, Roosevelt Rd, Taipei 106319, Taiwan, R.O.C %Tang
    \and Institut für Theoretische Physik und Astrophysik, Christian-Albrechts-Universität zu Kiel, Leibnizstraße 15, 24118 Kiel, Germany  %Wolf
   }
   \date{\today}

  \abstract
   {The dust and gas temperature in proto-planetary disks play critical roles in determining
   their chemical evolution and influencing planet formation processes.}
   {We attempted an accurate measurement of the dust and CO temperature profile in the edge-on disk of the Flying Saucer.}
   {We used the unique properties of the Flying Saucer, its edge-on geometry and its fortunate position in front of CO clouds with different brightness temperatures to provide independent constraints on the dust temperature. We compared it with the dust temperature derived using the radiative transfer code \textsc{DiskFit} and the CO gas temperature.}
   {We find clear evidence for a substantial gas temperature vertical gradient, with a cold (10\,K) disk mid-plane and a warmer CO layer where $T(r) \approx 27\,(r/100\,\mathrm{au})^{-0.3}$\,K. 
   Direct evidence for CO depletion in the mid-plane, below about 1 scale height, is also found. 
   At this height, the gas temperature is 15-20\,K, consistent with the expected CO freeze out temperature. 
   The dust disk appears optically thin at 345\,GHz, and exhibits moderate settling.
   }
  {}


   \maketitle
%
%________________________________________________________________
%
\section{Introduction}
\label{sec:intro}
Even if dust only represents $\sim$ 1\% of the total mass of molecular clouds, it is a key component in protoplanetary disks where planets form. Grain growth and dust coagulation is an important process to built planetary embryos and planets \citep{Birnstiel+2024} while gas-grain interactions in the cold part of protoplanetary disks allow a rich chemistry \citep{Oberg+2023,Gavino+etal_2021}. 
Hence, understanding grain composition and size, density and thermal structures of dust disks are necessary steps to constrain planet formation. 

In this domain, ALMA has revolutionized our views of planet formation, thanks to its resolving power and sensitivity. Nowadays, observers begin to unveil the disk physics and chemistry at the proper scale to constrain planetary formation. This is particularly true for dust disks orbiting around T\,Tauri stars. ALMA has shown how dust disks can be disturbed under the action of planet formation which generates gaps, rings and spirals gravitationally linked to young embedded planets (e.g. ALMA large program DSHARP, \cite{Andrews+2018}). Our understanding of the dust evolution has also improved, NOEMA,  ALMA and VLA observations revealing evidences for grain size variations with radius \citep{Guilloteau+etal_2011,Perez+2012,Tazzari+etal_2016,Tazzari+etal_2021,Guidi+etal_2022}. High angular images of dust disks have also shown that large grains in Class II disks have already settled onto disk mid-planes  \citep[e.g.][]{Guilloteau+2016,Villenave+etal_2020,Villenave+etal_2022}, as it was expected from hydrodynamical simulations \citep{Fromang+etal_2009}. Infrared observations obtained with the JWST currently unveil the vertical dust stratification \citep{Duchene+2024}. 

This paper focus on the thermal structure. Determining the thermal distribution in dust disks remains difficult even taking into account the improvements mentioned above, because it requires dedicated sensitive multi-wavelength observations and detailed models \citep{Ueda+2023}. 
We use a completely different approach based on dust and CO observations of the edge-on disk of the Flying Saucer \citep{Grosso+2003} which is seen in silhouette against several bright molecular clouds.
The method was originally developed by \cite{Guilloteau+2016} who applied it to $0.5''$ angular resolution data of the Flying Saucer disk to estimate the dust temperature of the mid-plane. 
We use here ALMA project 2023.1.00907.S to improve this study at higher angular resolution, 22 au at the $\rho$ Oph distance of 120\,pc, compared to 70\,au for the previous data set. 
We also measure the CO gas temperature inside the whole disk using the CO 2-1 emission line. 
We also introduce another direct method to measure both the CO gas and dust temperature. We then compare these direct methods with a simple classic modeling performed using the radiative transfer code \textsc{DiskFit}. 

\section{Observations and Imaging}
\label{sec:observation}
The ALMA project 2023.1.00907.S was observed in 2024.  
It covers two spectral setups, one in Band 6 and one in Band 7, to observe a maximum number of spectral lines at a spatial resolution as high as 0.18$''$ and spectral resolutions ranging from 40 m/s to 0.2 km/s, depending on the spectral line. We focus here on the study of the thermal structure using CO and dust. Dutrey et al (in prep.) 
present the analysis of the spectral lines and discuss evidences for a common molecular layer. 
We also used archival data from Projects 2013.1.00387.S (P.I. S.Guilloteau) and ALMA 2013.2.00163.S (P.I. M.Simon) that offered continuum images at slightly different frequencies
but with lower angular resolution, about $0.5''$ \citep{Guilloteau+2016,Dutrey+etal_2017,Simon+2019}. 
Data reduction, which requires accurate proper motion corrections to allow a common analysis, is presented in the Appendix \ref{app:A-data}. Fig.\ref{fig:index} shows the total flux as a function of frequencies derived from all continuum images. 

\begin{figure}
\centering
\includegraphics[width=0.85\columnwidth]{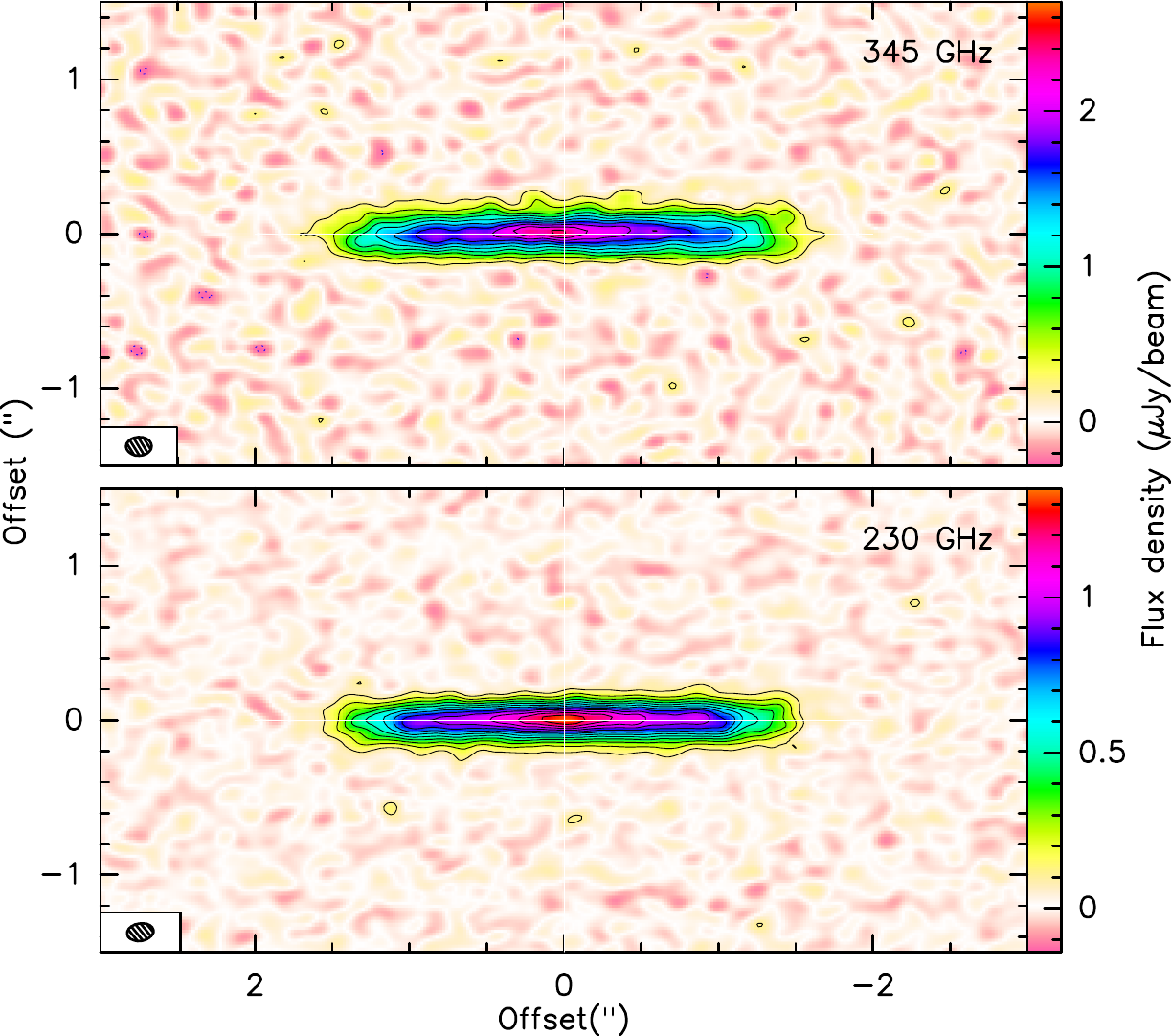}
\caption{High resolution images of the continuum emission from the Flying Saucer disk at 230 and 345 GHz. Contour levels are in steps of 4 $\sigma$ (0.16 and 0.13\,K respectively). The images were rotated by 3.1$^\circ$ clockwise to align the
disk major axis along the X axis. }
\label{fig:cont}
\end{figure}

The new continuum images are shown in Fig.\ref{fig:cont}. The angular resolution is $0.173\times0.114''$ at PA\,$98^\circ$ at 230 GHz, and $0.165\times0.122''$ at PA\,$91^\circ$ at 345 GHz respectively. Fig.\ref{fig:radial} shows the brightness profile along the disk mid-plane at the two frequencies. The new high angular resolution $^{12}$CO 2-1 channel maps are shown in Fig.\ref{fig:channels}. 

\section{Analysis and Modeling}
\label{sec:analysis}
\subsection{Direct analysis of dust emission}
\label{sub:direct}
Fig.\ref{fig:index} reveals an apparent spectral index below 2. This is suggestive of relatively low kinetic temperature, for which the deviations from the Rayleigh Jeans
regime are larger at 345 GHz than at 230 GHz. 
In the uniform slab approximation, we find a $T_b(230)/T_b(345)$ brightness ratio of 1.25 in
the 0-140 au range (see Fig.\ref{fig:radial}). Although such a ratio can be obtained from a black body emission at 13\,K, with the moderate dust opacities (0.2-0.5), lower temperatures are required because for all reasonable dust properties the opacity is larger at 345 GHz than at 230 GHz (see Appendix \ref{app:B-temperature}). 

\begin{table*}[!ht]
\caption{Dust and CO disk modeling results.}
\centering
\begin{tabular}{lrclllrcl}
\hline
Symbol & \multicolumn{3}{c}{From dust} & Unit & Parameter &  \multicolumn{3}{c}{From CO}   \\
\hline
$\beta_0$ & $0.7$ & $\pm$ & $0.3$ & & Dust opacity index at 100\,au\\
$\beta_r$ & $0.5$ & $\pm$ & $0.4$ & & Radial opacity index slope \\
$\Sigma_0$ & $1.7\,10^{23}$ & $\pm$ & $0.4\,10^{23}$  & cm$^{-2}$ & H$_2$ surface density at 100\,au \\
$\Sigma_g$ & 0.71 & $\pm$ & 0.04 & g\,cm$^{-2}$ & Surface density & $1.7\,10^{17}$ & $\pm$ &
$0.1\,10^{17}$ cm$^{-2}$  \\
$p$        &  $0.12$ & $\pm$ & $0.06$  &    & Surface density exponent &  1.35 & $\pm$ & 0.05 \\
$T_0$      & $9.0$ & $\pm$ & $1.0$   & K  & Mid-plane Temperature at 100\,au & 9.2 & $\pm$ & 0.8 \\
$q$        & $0.09$ & $\pm$ & $0.09$ & & Temperature exponent & -0.2 & $\pm$ & 0.2 \\
$T_\mathrm{atm}$ &  & & & & Atmosphere Temperature & 26.6 & $\pm$ & 0.1 \\
$q_\mathrm{atm}$ & & & & & Exponent & 0.28 & $\pm$ & 0.01 \\
$\delta$ & & & & & Gradient steepness & 1.0 & $\pm$ & 0.2 \\
$z_q$ & & & & & Amospheric Height & 1.66 & $\pm$ &  0.03 $H(r)$ \\  
$H_0$      &  $8.6$ &  $\pm$ & $0.35$  & au & Scale height at 100 au  & 19.1 & $\pm$ & 0.3\\
$h$        &  -0.86 & $\pm$ & 0.07  &    & exponent of scale height  & -0.9 & $\pm$ & 0.1 \\
$z_d$  & & & & & Depletion Height & 0.7 & $\pm$ &  0.2 $H(r)$ \\ 
$R_\mathrm{out}$ &  $187$ &  $\pm$ &  $2$ & au & Outer radius & 300 & $\pm$ & 4 \\
$R_\mathrm{int}$ &  $ < 2$ & &  & au & Inner radius & 24 & $\pm$ & 1 \\
\hline
\end{tabular}
  \label{tab:disk}
  \tablefoot{$\Sigma(r)=\Sigma_0(r/100\,\mathrm{au})^{-p}$,
  $H(r) = H_0 (r/100\,\mathrm{au})^{-h}$, and $T(r) = T_0 (r/100\,\mathrm{au})^{-q}$.
   Errors are 1 $\sigma$.} 
\end{table*}

\subsection{The dust disk shadow onto the CO extended emission}
The Flying Saucer lies in front of several molecular clouds, which emits in CO at different velocities and with different intensities \citep{Guilloteau+2016}, creating a disk shadow clearly seen in the new high angular resolution $^{12}$CO 2-1 channel maps (Fig.\ref{fig:channels}).  
We used the ``shadow'' method developed by \citet{Guilloteau+2016} (see details in Appendix\,\ref{app:D-Model}) to repeat their analysis at our much improved angular resolution. 
A nearly constant temperature of $\sim 5.5$\,K is obtained across the disk, except towards the disk center where it rises up to 10\,K in the inner 20 au.
\begin{figure}
\includegraphics[width=\columnwidth]{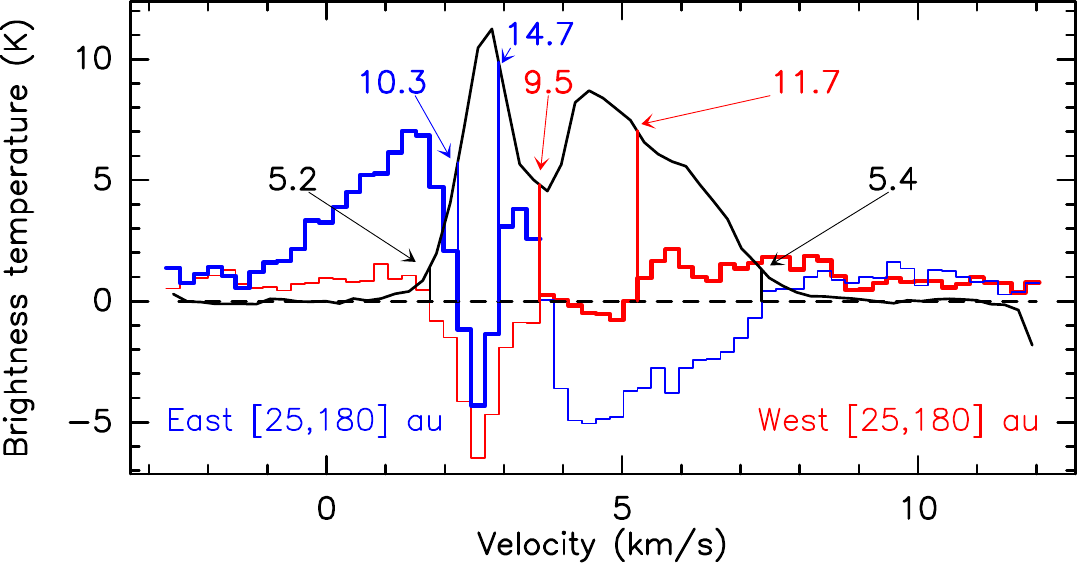}
\caption{Dust and CO temperature derivation. Numbers indicate the intrinsic temperatures
corresponding to the CO cloud brightness temperature at the positions of nulls
in the East (blue) and West (red) average mid-plane spectra. Black values are those
obtained for the dust.}
\label{fig:midplane}
\end{figure}

Another way to derive the temperature is to note that when the dust brightness
matches that of the background at a specific velocity $v$, 
$J_\nu(T_\mathrm{dust}) = J_\mathrm{cloud}(v)+J_{bg}$, the left-over signal $J_\mathrm{line}(v)$ vanishes. 
From the position-velocity map in Fig.\ref{fig:composite}, we compute the average spectra from the disk mid-plane between 24 and 180 au, and -24 and -180 au, displayed in Fig.\ref{fig:midplane} (in blue and red, respectively) with the background CO spectrum from the 30-m (in black). 
Whenever the computed spectra are 0, the dust (brightness) temperature is directly given by the CO spectrum intensity. 
Using the part of the spectra which are not contaminated by CO emission from the disk (low velocities of red spectrum, and high velocities of the blue one), and after correction from the cosmic background and Rayleigh Jeans deviation, we find dust temperatures of $5.2\,$K and $5.4$\,K. These mean temperatures are quite consistent with the median value derived before. From these ``shadow'' methods, we thus conservatively estimate that the dust temperature is roughly constant between 20\,au and the disk edge, 180\,au, with a mean value of $5.5\pm0.5$\,K

\subsection{Gas temperature determined from the background}
There are (four) other zeros in the two disk mid-plane spectra of Fig.\ref{fig:midplane}, at velocities where CO emission from the disk occurs. In principle, this can provide a measurement of the CO gas temperature. However, this cannot be done like for dust because of a lack of 
external reference (CO without any background). 
The analysis, whose details and limitations are presented in Appendix \ref{app:gas-temp}, indicates that the absorption of the CO background happens in gas with (mean) temperatures in the range 10 - 15\,K, since CO is likely to be thermalized.

\subsection{Dust temperature from disk model fitting}
\label{sub:dust} 
To obtain an independent estimate of the dust temperature, we fitted the continuum emission using a simple inclined disk model, using the ray tracing DiskFit tool \citep{Pietu+2007}. We followed the method explained in \citet{Guilloteau+etal_2011}, using a simple power law distribution for the dust surface density and temperature, radially truncated at an inner and outer radii. The disk is flared, with a Gaussian distribution in density as a function of distance to the mid-plane, and its scale height is a power law of the radius. The dust temperature only depends on radius.

We fit simultaneously the 230 and 345 GHz data assuming the dust opacity is given by 
$ \kappa(\nu) = \kappa_0 (\nu/\nu_0)^{\beta(r)}$
with the exponent following the prescription of \citet{Guilloteau+etal_2011}
$\beta(r) = \beta_0 + \beta_r\log(r/R_0)$, with $\kappa(\mathrm{230\,GHz}) = 3.5$\,cm$^2$ per gram of dust. 
We derive the H$_2$ surface density assuming a dust to gas ratio of 100.
We use a modified Levenberg-Marquardt minimization method, with several starting parameter sets to avoid local minima. 
Error bars were derived from the covariance matrix. 
Model parameters, best fit results and errors are given in Table\,\ref{tab:disk} (see also Appendix \ref{app:D-Model}).
Most parameters are well constrained, although only an upper limit for the inner radius could be found. 
Using this best fit model yields a typical optical depth of 0.1-0.5 across the disk (see Appendix \ref{app:C-opacity}). 

With only 2 observing frequencies, the error bar on the temperature remains significant because of the degeneracy with the opacity law \citep{Bohler+2013}. 
With our adopted opacity, the total disk mass is about $0.009 \Msun$.

\subsection{Gas temperature from CO disk model fitting}
\label{sub:co}
To represent the CO emission, we assume a disk model that has a vertical
temperature gradient \citep{Dartois+etal_2003} following the prescription from \citet{Dutrey+etal_2017,Foucher+2025} (see Appendix \ref{app:D-Model}).
The scale height is assumed to be a power law $H(r) = H_0(r/R_0)^{-h}$, and to represent the lack of CO emission in the mid-plane (see Fig.\ref{fig:channels} and \ref{fig:composite}), we assume that CO is not present below some depletion scale $z_d$, i.e. for $z < z_d = Z_d H(r)$. The CO surface density follows a power law. 
The disk is truncated at an inner and outer radii $R_{int}$ and $R_{out}$. 

The radiative transfer is solved by the \textsc{DiskFit} code, using the 30-m CO spectrum from \citet{Guilloteau+2016}, shown in Fig.\ref{fig:midplane}, as a background for ray-tracing as a function of velocities. 
Parameters are adjusted by a minimization based on visibility comparison in the $uv$ plane to propagate noise induced errorbars, see \citet{Pietu+2007}. 
Results are given in Table \ref{tab:disk}. 
The surface density law has little impact on the other results (except for the outer radius, that weakly depends on it) because the core of the CO line remains optically thick in most of the emitting region.

\section{Discussion and perspectives}
\label{sec:discussion}

\subsection{Mid-plane temperature}
\label{sub:mid-plane}
The dust temperatures derived through the ``disk shadow'' are consistent with those of \citet{Guilloteau+2016}, although based on a totally independent high resolution data set, with a dust temperature of 5-6\,K. 
However, \textsc{DiskFit} yields a higher value, around 9\,K, with slightly larger errors due to the coupling between the assumed dust emissivity profiles (function of $r$ and $\nu$) and temperature.

The discrepancy may have several origins. 
On one hand, since the single dish spectrum is the same, the absolute temperature scale in the shadow methods remains affected by the same calibration uncertainty. 
Also, these shadow methods assume no scattering, while self-scattering by dust may reduce the apparent disk brightness \citep{Kataoka+2015}; however, the low derived optical depth ($\sim 0.3$, see Appendix \ref{app:C-opacity}) seems to exclude this effect.
Finally, the shadow methods also assume that the 30-m spectrum represents uniform emission behind the dust disk. 
Since the residuals from the best CO model clearly reveal significant structures at scales of 2-10$''$, see Fig.\ref{fig:residual}, these inhomogeneities may bias the derived dust temperature, especially if combined with spatial structures in the dust disk. 

On another hand, the \textsc{DiskFit} results for dust are independent of the CO background, but rely on prescription for the dust emissivity spatial distribution. 
Given the limited frequency span, this assumption should have little impact on the derived temperature. We thus consider the \textsc{DiskFit} results more reliable, and 
that our observations and analysis indicate low temperatures for the dust, around 9\,K, throughout most of the disk. 
This value agrees remarkably well with the estimated gas temperature in the disk mid-plane. 
This is to be expected because at the mid-plane densities (given by $\Sigma_0/(\sqrt{\pi} H_0)$ from Table \ref{tab:disk}, 1 to a few $10^8$\,cm$^{-3}$), the dust imposes the gas temperature \citep{Chiang+1997}.

Concerning the flat temperature profile of the mid-plane (except within the inner 20\,au),
we note that non monotonic temperature profiles may result from
the frequency dependence of dust opacity that is controlled by 
optical properties of grains and grain growth \citep[e.g.][their Fig.7]{Isella+2009}.
These profiles may lead to an apparently flat mid-plane temperature when fitted by a simple power law. Such hidden gradients
could play a role in explaining the discrepancy found between \textsc{DiskFit} results and the analysis based on CO background.

\subsection{Gas disk and CO depletion}
The CO data indicate a substantial temperature gradient between the mid-plane, with a roughly constant temperature of 10\,K, and the upper layers where CO reaches temperatures decreasing from $\sim$\,35\,K near a radius of 50\,au to 15\,K at the disk edge. 
Note however that  because CO appears depleted in the mid-plane, the mid-plane temperature is essentially a result of the extrapolation of prescribed shape of the vertical temperature gradient.  While evidence for temperature gradients based on the
CO isotopologues are numerous \citep[e.g.][]{Rosenfeld+2013}, measurements of the mid-plane temperature remain rare. \citet{Dullemond+2020} measured 
a mid-plane temperature of 17\,K in HD\,163296 (based on optically thick CO), but were unable to conclude about CO depletion in this warmer source.

In previous studies, evidence for CO depletion came from observations of the optically
thin emission from isotopologues that allowed to locate radially the CO ``snowline''.
The morphology of CO emission in IM\,Lup \citep{Pinte+2018} is also suggestive of
partial CO freeze-out near the mid-plane.  Our study of the colder, edge-on disk of the Flying Saucer provide a first quantitative estimate of the thickness of
the depleted layer. 
The best fit model indicates strong CO depletion below about $0.7\,H(r)$
(at least a factor 500 to ensure low optical depth for the 2-1 transition in the mid-plane). 
Using the derived temperature profile, the temperature at this height varies
between 19\,K at 100 au to 15\,K at 300 au. 
This is in good agreement with the freeze out temperature of CO on pure ice 
\citep{Bisschop+2006,Minissale+2022}, although we used a prescribed shape for the depleted layer that did not incorporate any a priori temperature information (see App.\ref{app:D-Model}). 
It seems to exclude significant CO to CO$_2$ conversion on grains, that could raise the apparent freeze-out temperature up to 35\,K \citep[e.g.][]{Reboussin+2015,Bosman+2018,Ruaud+2022}. 
This latter mechanism may contribute to explain the higher freeze-out temperatures derived for DM\,Tau by \citet{Qi+2019} and TW\,Hya by \citet{Zhang+2017}.

\subsection{Dust Settling}
\label{sub:settling}
\citet{Bohler+2013} showed that settling may be diagnosed even by fitting a non-settled disk model as done in Sec.\ref{sub:dust}. In such a case, ignoring settling
may affect the derived disk parameters, e.g. the surface density profile, or the flaring exponent $-h$ that is underestimated. The only strong indicator
is the apparent scale height. In our case, the dust scale height (9\,au) is not
significantly lower than the predicted gas scale height (11\,au) if the mid-plane
temperature is 10\,K (at 100\,au). However, the flaring index is $-h = 0.9$ for
dust compared to the expected value of $1.5$ for the gas under hydrostatic equilibrium at 
constant mid-plane temperature, suggesting larger settling in the outer, less dense, regions. 

Instead of settling, the low flaring index from dust may suggest a self-shadowed disk leading to low mid-plane temperatures \cite[e.g.][]{Woitke+etal_2016}. However, a simple shadowing would result in a steep radial temperature gradient like in GG\,Tau \citet{Dutrey+2014}, in contrast to our flat temperature profile.  
Furthermore, the NIR images from \citet{Grosso+2003} indicate a flared morphology 
consistent with a flaring index of 1.25. 
Note here that the flaring index derived from CO only reflects the location of the bulk of CO emission, and not necessarily the underlying (H$_2$ and He) gas distribution.

The relatively moderate settling may suggest that the Flying Saucer host star is
still relatively young, as settling is less efficient in Class I sources \citep[e.g.][]{Villenave+etal_2023}. Unfortunately, the stellar properties are not sufficiently known
to constrain the system age.

In summary, our study, based on high angular resolution data, clearly shows the presence of an important vertical temperature gradient, the gas and dust mid-plane temperature being of the same order of 10\,K while the molecular layer (at about 1.5-2 hydrostatic scale heights) has a temperature derived from CO \textsc{DiskFit} modeling of 17--23\,K. These data also reveal direct evidence for strong CO depletion in the disk mid-plane, up to 0.7 CO scale heights where the
temperature reaches 16-18\,K. This is consistent with CO freeze-out
on grains, but not with significant conversion of CO to CO$_2$ on grain surfaces.

\begin{acknowledgements}
This work was supported by ``Programme National de Physique Stellaire'' (PNPS) and ``Programme National de Physique Chimie du Milieu Interstellaire'' (PCMI) from INSU/CNRS.
This work was partly supported by the Italian Ministero dell Istruzione, Università e Ricerca through the grant Progetti Premiali 2012 – iALMA (CUP C52I13000140001). This project has received funding from the European Union’s Horizon 2020 research and innovation programme via the European Research Council (ERC) Synergy Grant ECOGAL (grant 855130).
Y.W.T. acknowledges support through NSTC grant 111-2112-M-001-064- and 112-2112-M-001-066-.
This work was also supported by the NKFIH NKKP grant ADVANCED 149943 and the NKFIH excellence grant TKP2021-NKTA-64. 
This paper makes use of the following ALMA data: ADS/JAO.ALMA\#2013.1.00163.S, ADS/JAO.ALMA\#2013.1.00387.S and ADS/JAO.ALMA\#2023.1.00907.S. ALMA is a partnership of ESO (representing its member states), NSF (USA) and NINS (Japan), together with NRC (Canada), NSTC and ASIAA (Taiwan), and KASI (Republic of Korea), in cooperation with the Republic of Chile. The Joint ALMA Observatory is operated by ESO, AUI/NRAO and NAOJ.
\end{acknowledgements}

\bibliographystyle{aa}
\bibliography{ref-abaur.bib}

\begin{appendix}

\section{Data reduction}
\label{app:A-data}
Data were calibrated using the \textsc{Casa} calibration scripts provided by the ALMA observatory. 
Calibrated visibilities were then exported into UVFITS data format for imaging and further analysis, using the \texttt{IMAGER} package \footnote{see https://imager.oasu.u-bordeaux.fr}. 
Spectral resampling and conversion to the LSR velocity frame were done withing \textsc{Casa} prior to exportation to UVFITS format. 

\begin{figure}[!h]
\centering
\includegraphics[width=0.8\columnwidth]{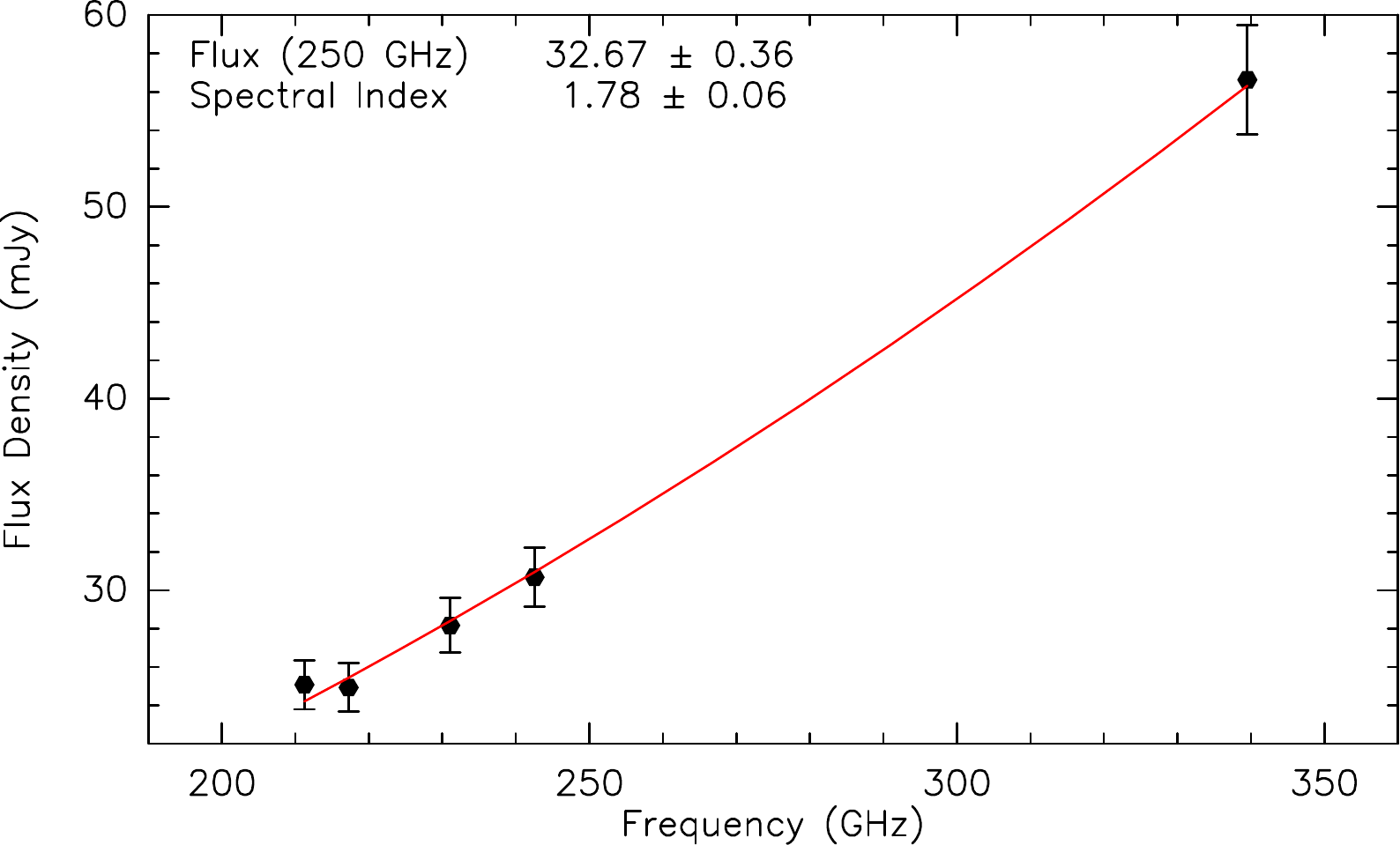}
\caption{SED of the Flying Saucer disk in 200-350 GHz range, with a power law fit superimposed.}
\label{fig:index}
\end{figure}

\paragraph{Proper motions:}
Having different epochs of observations spanning about 10 years, we were able to derive the proper motion of the source by comparing the apparent positions as a function of time, using the continuum data obtained for each data set. 
Our proper motion values,  $\mu_\alpha = -8.4\pm2.0$ and $\mu_\delta = -27.4 \pm  1.5$ mas/yr, are consistent with those derived for stars close to the Flying Saucer
in $\rho$ Oph (Elias 2-27, YLW 58, and YLW 18) using GAIA \citep{Gaia_DR3}. 
The distance of the Flying Saucer is unfortunately not known. From GAIA, Elias 2-27 is at 110\,pc, while YLW 58 and 18 lie at 135 and 137\,pc respectively. Since the Flying Saucer appears in front of molecular clouds, we adopt a distance of 120\,pc in this work. 

All observations were then registered to Epoch 2016.0 using the derived proper motion, the nominal position of the source being then R.A. 16:28:13.6979 and Declination  -24:31:39.491. 
Self-calibration using the continuum data was performed on the most compact configurations, but not for the most extended ones because the source structure (to first order a bar of apparent size $3 \times 0.3''$) is heavily resolved and does not leave enough flux on most baselines for this purpose. 
\begin{figure}[!h]
\centering
\includegraphics[width=0.9\columnwidth]{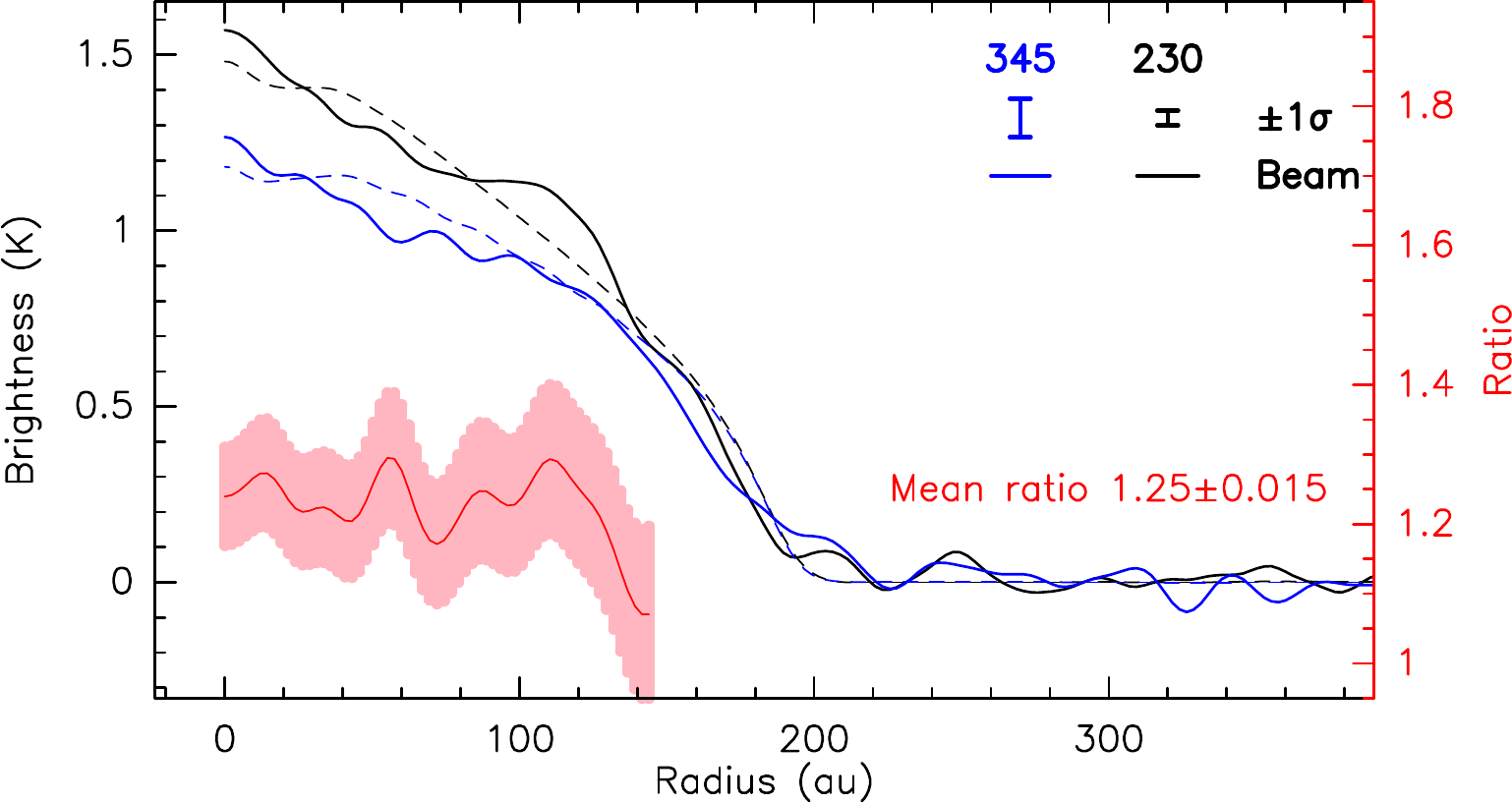}
\caption{Radial distribution of the brightness at 235  and 345 GHz along the disk mid-plane (continuous black and blue curves). Dashed lines indicate the best fit model prediction. The brightness ratio is given by the red curve, with $\pm 1 \sigma$ confidence interval in pink}.
\label{fig:radial}
\end{figure}

\paragraph{Flux calibration:}
Observations at different frequencies and epochs allowed us to derive the apparent spectral index of the continuum emission and check calibration consistency. 
The two configurations observed in each of Band 6 and Band 7 gave very consistent flux measurements (within $< 2$\,\%). 
However, the older observations from May 2015 (2013.1.00387.S and 2013.2.00163.S) were found to be too bright by about 5 and 7\,\% respectively, and were rescaled accordingly in flux before being merged with the newer data for imaging and analysis.

\begin{figure*}
\includegraphics[width=18.0cm]{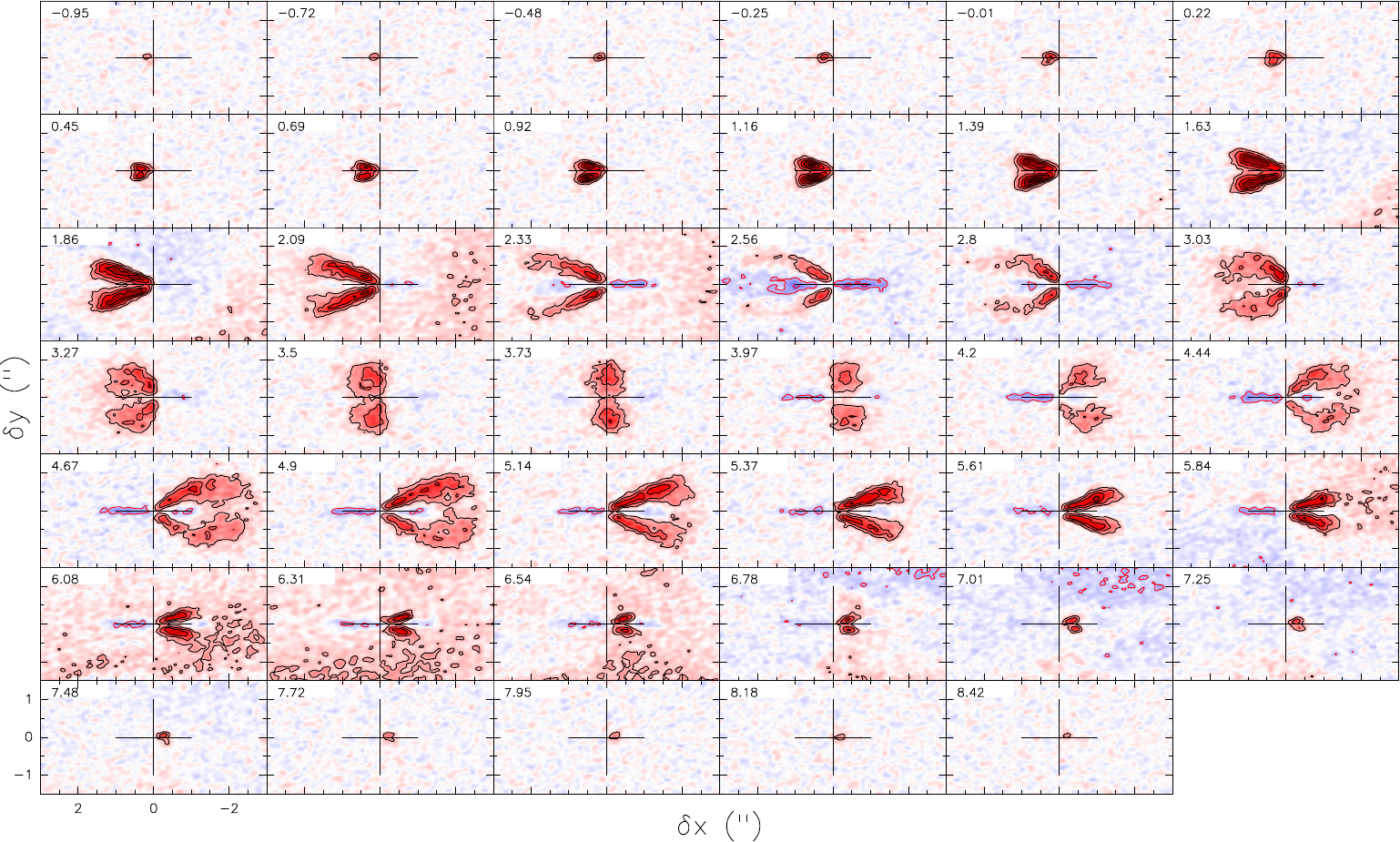}
\caption{Channel maps of CO emission.
Contour levels are in steps of 4.1\,K ($\simeq 5\sigma$).
Negative signal in blue, positive in red, with color scale
spanning -23 to 23\,K. Panels are labeled by the velocity, in range -0.95 to 8.42 
km\,s$^{-1}$.
\label{fig:channels}}
\end{figure*}
\begin{figure*}
\centering
    \includegraphics[width=17.0cm]{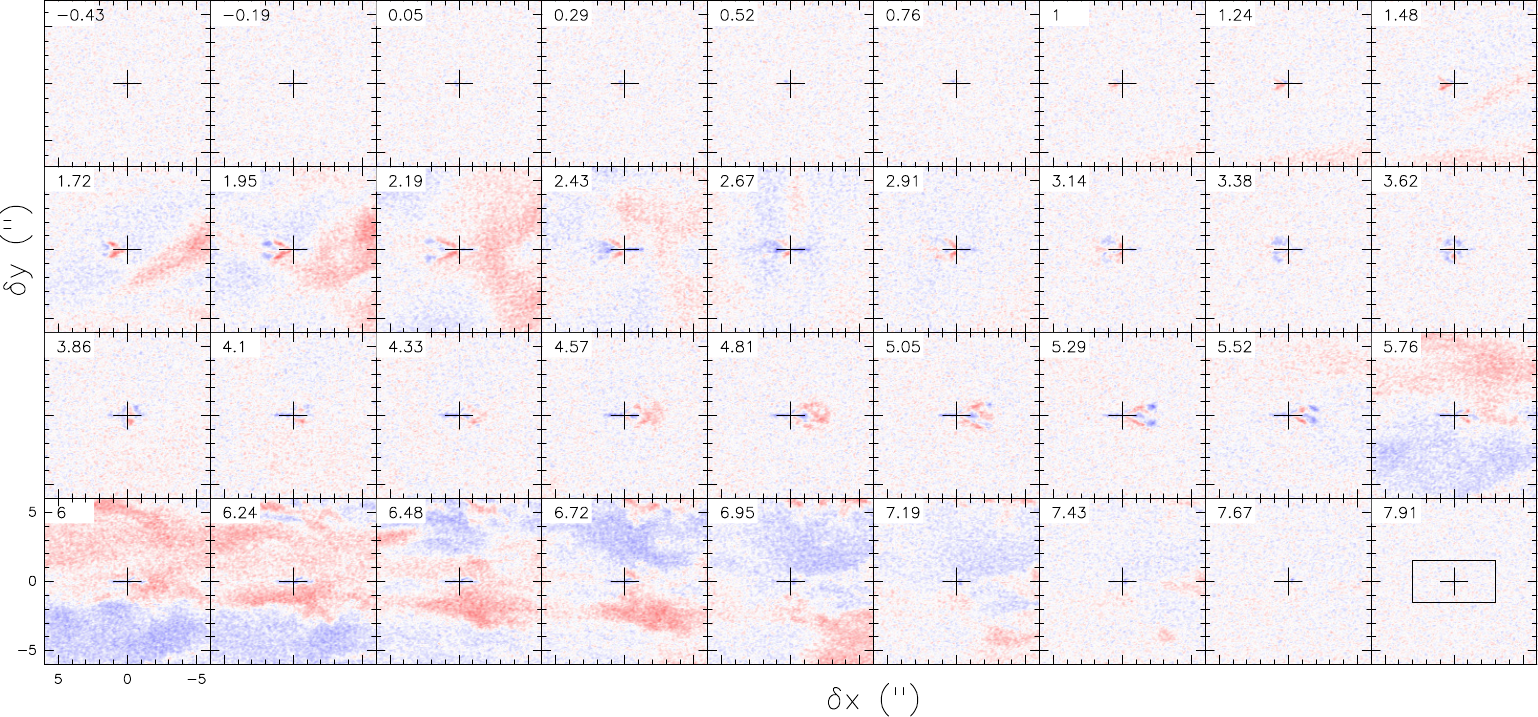}
\caption{12$''$ field channel maps of the residual emission after
model fitting. The color scale spans -23 to 23 K as in Fig\,\ref{fig:channels}, but data values in the residual only span the -9 to 9 K range. The box in the bottom right panel shows the area displayed in Fig.\ref{fig:channels}. Note the slightly different velocities compared to Fig.\ref{fig:channels}} 
\label{fig:residual}
\end{figure*}

\paragraph{Spectral and continuum imaging:}
Once the astrometric and flux calibration issues were solved, data from different configurations were combined together. 
All data was then imaged using a common grid,  with a pixel size of $0.025''$.
Various choices of robust weighting parameters were used in order to offer an adequate compromise between angular resolution, sensitivity and dirty beam shape (to minimize in particular the plateau of near sidelobes that occurs when combining two ALMA array configurations). 
A field of view of $25.6''$ was used for CO to best handle the extended background clouds (although the selected field of view had minor impact on the reconstructed image of the disk itself). 
The CO synthesized beam is $0.217 \times 0.144''$ at PA $104^\circ$, and the noise
level is $\sim 0.1$\,K at 0.23 km\,s$^{-1}$ spectral resolution.

The continuum images were produced after filtering the spectral line emission using the \texttt{UV\_FILTER} command of \texttt{IMAGER}. 
Bandwidth synthesis was used to obtain the images.
The resolution is $0.172\times  0.113''$ at PA $100^\circ$ at 230\,GHz, and $0.166 \times 0.121''$ at PA $90^\circ$ at 345\,GHz.
We tested several deconvolution methods, all giving consistent results within the expected noise level for the spectral line observations.  
The continuum data is however dynamic range limited, with actual noise level 1.5 to 2 times larger than the thermal noise, and dynamic range about 60 at high resolution.
All spectral and continuum images were finally rotated by 3.1$^\circ$ clockwise to align the disk major axis along the X axis. 
Figure \ref{fig:channels} presents the CO channel maps.

\FloatBarrier

\section{Temperature derivation}

\begin{figure}[!t]
\includegraphics[width=\columnwidth]{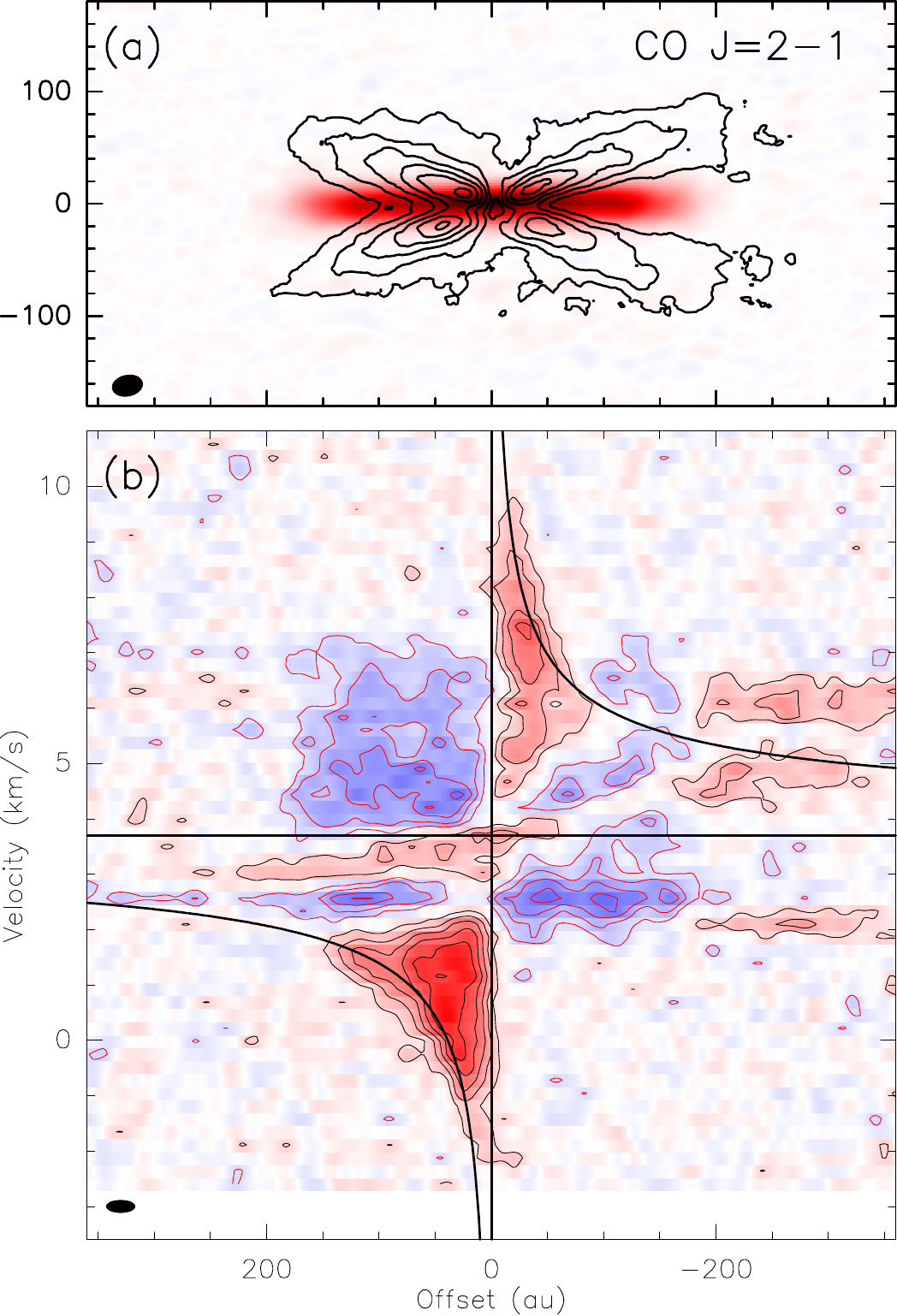}
\caption{
(a) Integrated CO line flux; contour spacing is 10 mJy/beam.km/s (7 K.km/s), overlaid on the continuum image (peak brightness is 2\,K). 
b) Position-Velocity diagram across the disk plane. Contour spacing is 2\,K, about $2.4\,\sigma$.
The black curve is the Keplerian velocity for a $0.6\,\Msun$ star. 
Negative signal in blue, positive in red.
}
\label{fig:composite}
\end{figure}

\label{app:B-temperature}
\begin{figure}
\includegraphics[width=\columnwidth]{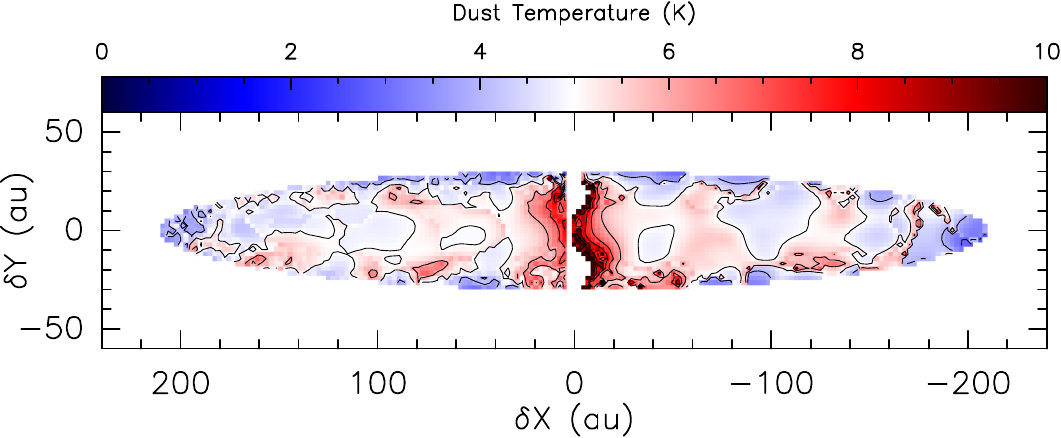}
\caption{Mean dust temperature distribution across the disk.
Contour step is 1\,K.}
\label{fig:temperature}
\end{figure}

\subsection{Direct analysis of dust emission}
\label{app:direct}

In a uniform slab, the brightness $T_b$ is given by
\begin{equation}
\label{eq:basic-rt}
T_b(\nu)= f \left( 1-\exp(-\tau(\nu)) \right) \left(J_\nu(T_d) - J_\nu(T_{bg})\right)
\end{equation}
where $J_\nu$ is the radiation temperature, i.e. the Planck function multiplied by $c^2/2k\nu^2$:
\begin{equation}
J_\nu(T) = \frac{h \nu}{k} \frac{1}{\exp(h\nu/(kT))-1} % \approx T - \frac{h\nu}{2k}
\end{equation}
$T_d$ the dust temperature, $T_{bg} = 2.7$ K the cosmic background temperature, $\tau$ the optical depth and $f$ a beam filling factor $<1$. 
The brightness ratio $S_{23} = T_b(235\,\mathrm{GHz})/T_b(345\,\mathrm{GHz})$ can be used to obtain an estimate to $T_d$. 
Given the similar geometrical thickness and spatial resolution at both frequencies (see Fig.\ref{fig:filling}), a common beam filling factor can be used. 
The uniform slab approximation is reasonable, since $S_{23}$ is approximately constant at 1.25 between 20 and 120 au (Fig.\ref{fig:radial}). 
In the optically thick limit, or when opacities are equal at both frequencies, this $S_{23}$ value corresponds to $T_d = 13$\,K. 
However, since for all reasonable dust properties the opacity is larger at 345 GHz than at 230 GHz (their ratio being 1.3 using $\beta$ from Table \ref{tab:disk}), this value is an upper limit: a partially optically thin medium would need to be colder.

\subsection{The shadow method for dust}
\label{app:shadow}
As discussed by \citet{Guilloteau+2016}, the Flying Saucer lies in front of several molecular clouds, which emits in CO at different velocities and with different intensities.  The dust disk appears in shadow against this bright, frequency variable background, providing a (so far unique) opportunity to measure the dust temperature in a totally independent way compared to usual methods. The disk shadow can be clearly seen in the high angular resolution $^{12}$CO 2-1 channel map shown in Fig.\ref{fig:channels}. 

The ``shadow'' method derives from the basic radiative transfer equation (Eq.\ref{eq:basic-rt})
that gives the flux (expressed in brightness) $J_m$ emerging from a medium of temperature $T_m$ as a function of velocity $v$,
\begin{equation}
J_m(v) = f \left(1-\exp(-\tau(v)) \right) \left(J_\nu(T_\mathrm{m}(v)) - J_{back}(v) \right)
\end{equation}
$J_\mathrm{back}$ is here the brightness temperature of any spatially extended
background, that is removed by the On-Off observing technique with single-dish, or filtered out by interferometers.

Far from the cloud velocities, $J_{back} = J_{bg} = J_\nu(T_{bg}) = 0.19$ K at 230 GHz is due to the cosmological background.
At velocity $v$, the dust disk is in front of an extended CO cloud whose apparent brightness, as measured by the IRAM 30-m telescope, is thus
\begin{equation}
J_\mathrm{cloud}(v) =  (1-\exp(-\tau_\mathrm{cloud} (v))) (J_\nu(T_\mathrm{cloud}(v)) - J_{bg})
\label{eq:cloud}
\end{equation}
so that of the dust disk brightness temperature at velocity $v$ becomes
\begin{equation}
\label{eq:zero}
J_\mathrm{line}(v) = f (1-\exp(-\tau)) (J_\nu(T_\mathrm{dust}) - J_\mathrm{cloud}(v) - J_{bg})
\end{equation}
We call $J_\mathrm{disk}$ the disk continuum brightness temperature
away from the cloud velocities. 
Subtracting the $J_\mathrm{disk}$ and $J_\mathrm{line}$ yields the optical depth and beam filling factor product
\begin{equation}
 f (1-\exp(-\tau)) = (J_\mathrm{disk} - J_\mathrm{line}(v))\, /\, J_\mathrm{cloud}(v)
\label{eq:fill}
\end{equation}
that we use to eliminate from Eq.\Ref{eq:zero}, and after some
re-arrangement we obtain
\begin{eqnarray}
J_\nu(T_\mathrm{dust}) & = &  
\frac{ J_\mathrm{line} J_\mathrm{cloud}(v)}{J_\mathrm{disk} -J_\mathrm{line}(v)} + J_\mathrm{cloud}(v) + J_{bg} \\
& = & 
\frac{ J_\mathrm{disk} J_\mathrm{cloud}(v)}{ J_\mathrm{disk} - J_\mathrm{line}(v) } + J_{bg}
\label{app:eq:tdust}
\end{eqnarray}
A position-velocity map showing $J_\mathrm{line}(v)$ across the disk mid-plane is displayed in Fig.\ref{fig:composite}.
The method can be applied to any line of sight to recover the dust temperature across the disk. The cloud emission being velocity dependent provides several independent estimates for each direction, and we take the median value to obtain the dust temperature map displayed in Fig.\ref{fig:temperature}.
Compared to the \citet{Guilloteau+2016} data, the current data improve the spatial resolution
by a factor 2 along the radius and 3 in the vertical direction. 
A nearly constant temperature of $\sim 5.5$\,K is obtained across the disk, except towards the disk center where it rises up to 10\,K in the inner 20 au.

Another derivation is possible  when $J_\mathrm{line}(v) = 0$ in Eq.\ref{eq:zero}, but this only occurs at two velocities, (2.7 and 7.3 km\,s$^{-1}$, Fig.\ref{fig:midplane}) providing less redundancy in the determination of $T_\mathrm{dust}$.

\subsection{The shadow method for gas}
\label{app:gas-temp}
Gas around the disk mid-plane also appears in absorption against the CO clouds. However, contrary to dust, there is no reference emission flux ($J_\mathrm{disk}$) that we can use to eliminate the $f(1-\exp(-\tau))$ factor. 
So we are left with the methods relying only on the zero values of Eq.\ref{eq:zero} found at some velocities in Fig.\ref{fig:midplane}, where neither $f$ nor $\tau$ matters. This method has a number of limitations, however.

First, the spectrum in Fig.\ref{fig:midplane} being an average
over the disk radii, at velocity offsets $dV$ larger than 1.7\,km\,s$^{-1}$ from the systemic velocity (3.70\,km\,s$^{-1}$) emission only comes from regions smaller than  
$r(dV) = (G\,M_*)/dV^2  = 100\, (2.3/dV)^2$\,au 
due to the Keplerian rotation (see Fig.\ref{fig:composite}). Therefore, the mean CO brightness temperature is only valid up to this radius.
Fortunately, all zeros are within the [2.0,5.4]\,km\,s$^{-1}$ velocity range and do not need to be corrected.
Strictly speaking, a similar correction is also needed for emission near the systemic velocity, since it only extends up to 80-100 au because of the velocity projection, but this does not affect the position (in velocity) of the zero. 

Second, the disk may (or may not) be optically thick in CO, hiding (part of) the dust emission. However, this is a small effect on the apparent brightness (only 0.2\,K) that we can neglect.

Third, Eq.\ref{eq:zero} is for a homogeneous medium. In practice,
we only measure the average brightness temperature, and its value can be interpreted as being due to either cold gas filling the beam, or warmer gas with a smaller filling factor.  Assuming $f=1$, we get temperatures ranging from 9.5\,K to 14.7\,K. The temperature of 9.5\,K appearing in Fig.\ref{fig:midplane} at 3.7\,km\,s$^{-1}$ does not account for the apparent disk extent (caused by the velocity gradient) at this velocity. Using a more appropriate beam filling factor of 0.5 for this case raises the temperature to 14.2\,K.

\section{Dust optical depth}
\label{app:C-opacity}

\begin{figure}[!h]
\centering 
\includegraphics[width=0.8\columnwidth]{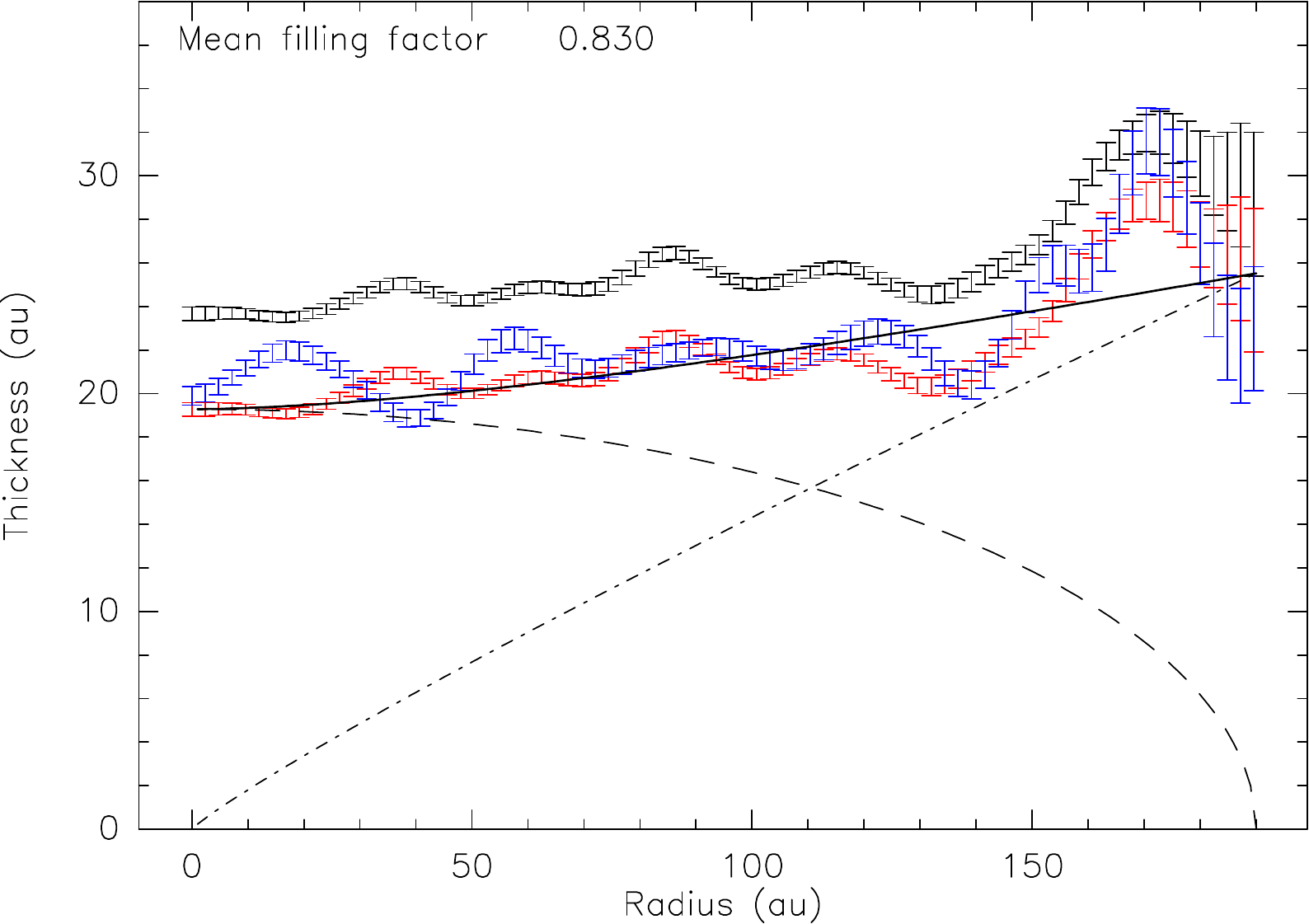}
\caption{Apparent (black) and intrinsic (red) thickness of the continuum emission at 230 GHz as a function of radius;  in blue, intrinsic thickness at 345 GHz.  The dash-dotted line indicates the width due to the best fit scale height,
the dashed line the expected width due to the disk inclination of $\sim 86^\circ$, and the solid line the resulting total width.}
\label{fig:filling}%
\end{figure}

\paragraph{Direct determination:}
We can estimate the dust optical depth by comparing the apparent brightness of the continuum to the temperature derived from the shadow, but this requires an estimate of the beam filling factor $f$.  
Since the disk is well resolved in radius, $f$ is controlled by the thickness. 
The intrinsic thickness is derived (to first order) by deconvolving the apparent thickness from the effective resolution, here the beam minor axis, as the beams are essentially elongated parallel to the radius. 
Apparent sizes were derived by Gaussian fit to the observed (vertical) profile. 
Apparent and corrected sizes are displayed as a function of radius in Fig.\ref{fig:filling}. 
The mean filling factor is $0.83$. 
It should be noted that this correction is an underestimate of the required one, because the images are dynamic range limited due to significant phase noise remaining on the long baselines. 
This results in a seeing effect that should be combined in quadrature with the beam size. 
This may bring down the beam filling factor to $0.7$, for which the peak brightness still only reaches $2.2$\,K from Fig.\ref{fig:radial}.
When compared to the estimated dust temperature, 9\,K, this indicates dust opacities of order 0.25.

\paragraph{From the model:}
While the \textsc{DiskFit} tool can produce optical depth maps for the best fit model parameters, a simple analytical verification is possible because, for dust, 
the best fit model is essentially one of an isothermal, uniform surface density disk. In this vertically isothermal dust disk, the mid-plane density is given by 
\begin{equation}
 \rho(r,z=0) = \Sigma_g / (\sqrt{\pi} H_0) (r/R_0)^{(-p+h)} \approx \rho_0 (R_0/r)    
\end{equation}
since $-p+h \approx -1$ (see Table\,\ref{tab:disk}, $R_0 = 100$\,au being the reference radius at which $\Sigma_g$ is given.

Assuming edge-on geometry, at an impact parameter $R$, the line of sight optical depth in the midplane is approximately given by 
\begin{equation}
    \tau(R,z=0,\nu) \approx 2 \kappa(\nu) \int_R^{R_{out}} \rho(r,z=0) dr
\end{equation}
(provided $R$ remains reasonably small compared to $R_{out}$, which
allows us to ignore the incidence angle at each radius, an approximation that is
valid up to about 120 au), i.e;
\begin{eqnarray}
\tau(R,\nu) & \approx & 2\,\kappa(\nu) \rho_0\,R_0\log(R_{out}/R) \\
            &  =      & 2\,\kappa(\nu) \Sigma_g R_0/(\sqrt(\pi) H_0) \log(R_{out}/R) 
\label{app:eq:odust}
\end{eqnarray}
that gives $   \tau(R,235\,\mathrm{GHz}) \approx 0.2 \log(R_{out}/R) $ using numbers from Table \ref{tab:disk}, yielding $\tau(70\,\mathrm{au},235\,\mathrm{GHz}) =0.2$ in the mid-plane.
The disk becomes optically thick only for radii smaller than 2 au, consistent with an inner radius upper limit found in the minimization process. In this edge-on assumption, the opacity decreases with the dust scale height.
The small tilt from edge-on view will slightly smear out the opacity distribution.

\section{Disk Model}
\label{app:D-Model}
In our disk model, the density distribution is assumed to be a Gaussian as a function of height $z$ above the mid-plane, \mbox{$n(r,z) = n_0 \exp(-(z/H(r))^2)$}, with a scale height being a power law, $H(r) = H_0 (r/R_0)^{-h}$, and the gas temperature distribution is
\begin{equation}
    \label{eq:temp}
    T(r,z) = \left(T_{mid}(r)-T_{atm}(r)\right)\left(\cos\left(\frac{\pi z}{2z_q H(r)}\right)\right)^{2\delta}+T_{atm}(r)
\end{equation}
for $z < z_q$, and $T_{atm}(r)$ above, with the ``atmosphere'' temperature given by: 
\begin{equation}
    \label{eq:tatm}
    T_{atm}(r) = T^{0}_{atm}\left(r/R_0\right)^{-q_{atm}}
\end{equation}
and the mid-plane temperature by:
\begin{equation}
    \label{eq:tmid}
    T_{mid}(r) = \mathrm{min}\left(T_{atm}(r),T_{0}\left(r/R_0\right)^{-q}\right)
\end{equation}
$z_q$ indicates the start of the (isothermal) atmosphere, $\delta$ the gradient
steepness. For $\delta \simeq 1$ the steepest gradient is in the transition zone between mid-plane and atmosphere at $z_q H(r)/2$. We do not impose the scale height to
be related to the mid-plane temperature. Instead, it is determined to 
represent the distribution of the CO molecules that best fit the observed data,
with the further assumption that no CO is present below a depletion height $z_d = Z_d H(r)$, where $Z_d$ is another fitted parameter in the model.

To reproduce the velocity pattern, we also account for the geometric and kinematics parameters: 
orientation ($PA=3.1\pm0.2^\circ$), inclination ($i=86.3\pm0.2^\circ$), systemic velocity ($V_\mathrm{LSR}=3.71\pm0.01$\,km\,s$^{-1}$), stellar mass (M$_*=0.59\pm0.01 \Msun$), and local line width ($\delta V = 0.20\pm0.01$\,km\,s$^{-1}$, assumed constant across the disk). 
These parameters are not significantly coupled to any of the others. With power laws
for surface density and scale heights, our model cannot represent the disk edge shape, and only yields results relevant to the 50\,au to 250\,au radius range.

Significant coupling exists between $T_\mathrm{mid}$, $Z_d$, $H_0$ and $\delta$. To alleviate this problem, we performed fits fixing one parameter and solving for the others. 
We also explored an alternate functional shape, that is equivalent to Eq.\ref{eq:temp} only if $\delta = 1$, namely
\begin{equation}
    \label{eq:tempsin}
    T(r,z) = \left(T_{atm}(r)-T_{mid}(r)\right)\left(\sin\left(\frac{\pi z}{2z_q H(r)}\right)\right)^{2\delta}+T_{mid}(r)
\end{equation}
All explored cases yielded low $T_\mathrm{mid}$, and the temperature at $z = z_d$ was always around 17\,K.

\end{appendix}
\end{document}